\documentclass{ws-acs}

\usepackage{amsmath,amsfonts,amssymb}
\usepackage{url}
\usepackage{graphicx}

\begin{document}

\makeatletter
\def\@biblabel#1{[#1]}
\makeatother

\markboth{Ihor Lubashevsky}{Fuzzy rationality and emergent phenomena}

%
\catchline{}{}{}{}{}
%

\title{DYNAMICAL TRAPS CAUSED BY FUZZY RATIONALITY\\AS A NEW EMERGENCE MECHANISM}

\author{\footnotesize IHOR LUBASHEVSKY}

\address{Division of Computer Science, University of Aizu\\
Aizu-Wakamatsu, Fukushima 965-8560, Japan\\
i-lubash@u-aizu.ac.jp}

\maketitle

\begin{history}
\received{(received date)}
\revised{(revised date)}
\end{history}

\begin{abstract}
A new emergence mechanism related to the human fuzzy rationality is considered. It assumes that individuals (operators) governing the dynamics of a certain system try to follow an optimal strategy in controlling its motion but fail to do this perfectly because similar strategies are indistinguishable for them. The main attention is focused on the systems where the optimal dynamics implies the stability of a certain equilibrium point in the corresponding phase space. In such systems the fuzzy rationality gives rise to some neighborhood of the equilibrium point, the region of dynamical traps, wherein each point is regarded as an equilibrium one by the operators. So when the system enters this region and while it is located in it, maybe for a long time, the operator control is suspended. To elucidate a question as to whether the dynamical traps on their own can cause emergent phenomena the stochastic factors are eliminated from consideration. In this case the system can leave the dynamical trap region only because of the mismatch between actions of different operators. By way of example, a chain of oscillators with dynamical traps is analyzed numerically. As demonstrated the dynamical traps do induce instability and complex behavior of such systems.
\end{abstract}


\section{Introduction}\label{sec:Int}

During the last decades there has been considerable progress in describing social systems based on physical notions and mathematical formalism developed in statistical physics and applied mathematics  (for a recent review see, e.g., \cite{GR1,GR2,GR3,GR4,GR5,GR6}). In particle, the notion of energy functional (Hamiltonian) and the corresponding master equation were employed to simulate opinion dynamics, the dynamics of culture and languages (e.g., \cite{GR3,GR6,OD1}); the social force model inheriting the basic concepts from Newtonian mechanics was used to simulate traffic flow, pedestrian motion, the motion of bird flocks, fish school, swarms of social insects (e.g., \cite{GR3,SFM1,SFM2}). A detailed review on other techniques based on kinetic theory, fluid dynamics, the Ginsburg-Landau equations, etc. applied to traffic flow and similar problems can be found also in Refs.~\cite{Helb1,Shc1,Nag1}. Continuing the list of examples, we note the application of the Lotka-Volterra model and the related reaction-diffusion systems to stock market, income distribution, population dynamics \cite{LVRD}. The replicator equations developed initially in the theory of species evolution were applied to the moral dynamics \cite{MD}. The notion of a fixed-point attractor as a stable equilibrium point in the system dynamics that corresponds to some local minimum in a certain potential relief is widely met in social psychology \cite{SocPsy}. The latter is extended even to collections of such fixed point attractors to form a basin. Besides, social psychology uses the notion of latent attractors (i.e. invisible ones under the equilibrium and whose presence affects strong perturbations), periodic attractors representing limit cycles, and deterministic chaos. In addition, the concept of synchronization of interacting oscillators was used to model social coordination \cite{SCD}.

In spite of these achievements we have to note that the mathematical theory of social systems is currently at its initial stage of development. Indeed, animate beings and objects of the inanimate world are highly different in their basic features, in particular, such notions as willingness, learning, prediction, motives for action, moral norms, personal and cultural values are just inapplicable to inanimate objects. This enables us to pose a question as to what \emph{individual} physical notions and mathematical formalism should be developed to describe social systems in addition to the available ones inherited from modern physics. For example, Kerner's hypothesis about the \emph{continuous} multitude of metastable states representing the synchronized phase of traffic flow, on one hand, stimulated developing the three-phase traffic model explaining a number of observed phenomena in congested traffic flow \cite{KE,KB}. On the other hand, a \emph{microscopic} mechanism enabling the coexistence of many different metastable states actually at the same point of the corresponding phase space  is up to know a challenging problem. 

The present paper discusses one of such notions, namely, the fuzzy rationality which is a specific implementation of the bounded capacity of human cognition \cite{FuzzyRat}. Its particular goal is to demonstrate that the fuzzy rationality can be responsible for complex emergent phenomena in social systems.   

The paper is organized as follows. Section~\ref{sec:FRGE} provides an explanation for the basic ideas and mathematical constructions to be used in formulating the main model in Sec.~\ref{sec:LBM}. Results of its numerical simulation and their discussion are presented in Sec.~\ref{sec:NM}. Section~\ref{sec:Concl} concludes the paper.

\section{Fuzzy rationality and governing equations}\label{sec:FRGE}

To elucidate the problem at hand, this section considers a simple imaginary system governed by an operator (individual), for example, a person driving  a car. The dynamics of the given system is represented as the motion of a point $\{x,y\}$ on a phase plane $\mathbb{R}_{xy}$. 

\subsubsection*{The limit of perfect rationality}

\begin{figure}[t]
\begin{center}
  \includegraphics[width=0.9\textwidth]{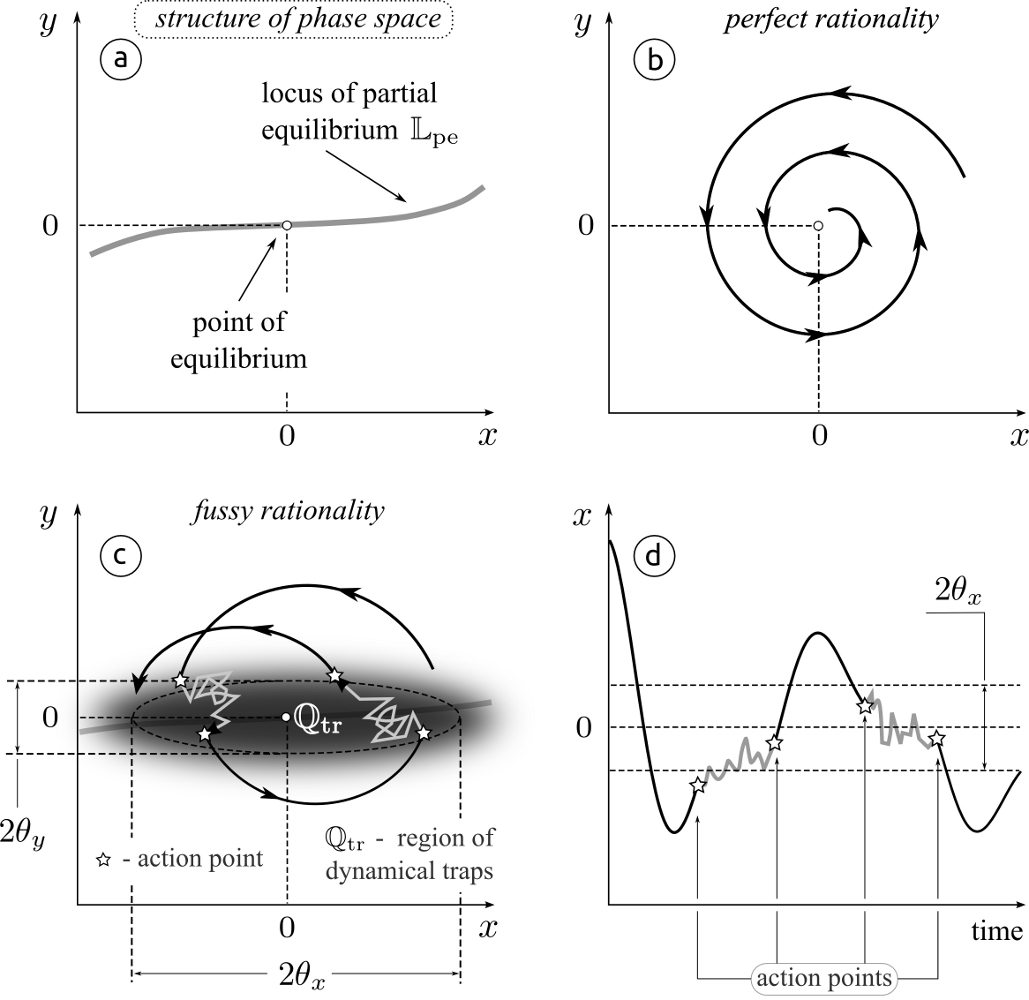}
\end{center}
\caption{The presumed structure of the phase space $\mathbb{R}_{xy}$ of the system considered in Sec.~\ref{sec:FRGE} (a); a schematic illustration of its dynamics near the stable equilibrium point $\{0,0\}$ in the cases of the perfect rationality (b) and the fuzzy rationality (c,d). Here, not to overload the drawings, the frame origin is shifted from the point $\{0,0\}$. }
\label{F1}
\end{figure}

We presume that if the operator was able to govern the system perfectly following a certain optimal strategy then its dynamics would be described by the coupled equations 
\begin{subequations}\label{Int:1}
\begin{align}
\label{Int:1x}
  \tau\frac{dx}{dt} &= F_x(x,y)\,,
\\   
\label{Int:1y}
  \tau\frac{dy}{dt} &= F_y(x,y)\,.
\end{align}
\end{subequations}
Here $\tau$ is a time scale characterizing the operator perception delay,  the ``forces'' $F_x(x,y)$ and $F_y(x,y)$ are determined by \emph{both} the physical regularities of the system mechanics and the active behavior of the operator in controlling the system motion. The origin $\{0,0\}$ of the coordinate frame is placed at the equilibrium point of system~\eqref{Int:1}, i.e., the equalities
\begin{align}\label{FR:2}
   F_x\big|_{\substack{x=0 \\ y=0}} & = 0\,, &  F_y\big|_{\substack{x=0 \\ y=0}} & = 0
\end{align}
are assumed to hold. In this context the perfect rationality of the operator means his ability to locate precisely the current position of the system on the phase plane $\mathbb{R}_{xy}$, to predict strictly its further motion, and, then, to correct the current motion continuously. Exactly in this case it is possible to consider that the operator orders the strategies of behavior according to their preference and then chooses the optimal one.
As a result the equilibrium point $\{0,0\}$  must be stable when the aim of operator actions is to keep the system in close vicinity to this point (Fig.~\ref{F1}a,b). 

The motion of the given system has been presumed to be a cumulative effect of the physical regularities and the operator actions. The notion of partial equilibrium implements this feature. Namely, the operator is considered to be able to halt the system motion at a certain multitude of points $\mathbb{L}_\text{pe}$ in the phase space $\mathbb{R}_{xy}$ to be called the locus of partial equilibrium and treated here as some smooth curve. So in the case of perfect rationality a gradual system motion towards the equilibrium point $\{0,0\}\in \mathbb{L}_\text{pe}$ is due to the intelligent actions of the operator which can locate this point on the plane $\mathbb{R}_{xy}$ precisely. The coordinate frame under consideration has been chosen in such a manner that the $x$-axis be tangent to the partial equilibrium locus $\mathbb{L}_\text{pe}$ at the point $\{0,0\}$ (Fig.~\ref{F1}a).       

Let us touch on the dynamics of a car following a lead car moving ahead with a fixed velocity $V$ to exemplify these constructions. The motion of the following car is usually described in terms of the headway distance $h$ and its velocity $v$ whose time variations are governed by the social force model generally written as
\begin{align}\label{eq:SFM}
 \frac{dh}{dt} &= V-v\,, & \frac{dv}{dt} & = F_v(v,h,V)\,, 
\end{align}
where $F_v(v,h,V)$ is a certain function.\footnote{Model~\eqref{eq:SFM} also admits a generalization relating the current acceleration $a(t)=dv/dt$ to the headway distance $h(t-\overline{\tau})$ and the car velocity $v(t-\overline{\tau})$ taken at the previous moment of time with some time shift $\overline{\tau}$ (for a review see, e.g., \cite{Helb1}).} 
In specifying the function $F_v(v,h,V)$ a driver is typically assumed to respond to the combined effect of two stimuli. One of them is to keep the speed of his car equal, on the average, to the lead car speed $V$ and can be quantified in terms of the relative velocity $u=v-V$. The other is to maintain an optimal headway distance $h_\text{opt}(v,V)$ generally determined by the values of $v$ and $V$. This stimulus similarly can be quantified by the difference $h-h_\text{opt}(v,V)$. So $\{h_\text{opt}, V\}$ is the equilibrium point of the car following in the phase space $\mathbb{R}_{hv}$ and the partial equilibrium locus $\mathbb{L}_\text{pe}$ is the line $v=V$. Indeed, by simple kinematic reasons any point on this line corresponds to a steady state arrangement of the two cars which can be frozen by the driver just fixing the car velocity, whereas keeping the headway distance equal to its optimal value $h_\text{opt}(V,V)$ is due to the driver intelligent action.

It is worthwhile to note that the social force model~\eqref{eq:SFM} matches actually the perfect rationality in driver behavior. In fact, the \emph{detailed} description of driver actions requires a certain extension of the phase space including, at least, the car acceleration $a$ as an individual phase variable. The matter is that, on one hand, by physical reasons the driver cannot change freely the position $x$ and velocity $v$ of his car, he is able to affect the car dynamics via  changing the acceleration $a$ only. On the other hand, the acceleration on its own contributes to the driver perception of the car motion quality. So in the approximation of perfect rationality the description of the car following is reduced to the problem of minimizing a certain cost functional whose integrand, a cost function, depends on the headway distance $h$, the velocity $v$, and also the acceleration $a$ \cite{me1}. As a result the governing equation  
contains $\ddot{a}$ in the leading order and, thus, does not meet the Newtonian mechanics paradigm. However, since the rational driver can perfectly predict the car motion, the final governing equation is reduced to the social force model~\eqref{eq:SFM}, where the ``force'' $F_v(v,h,V)$ depends not only on the current values of the headway distance $h$ and the car velocity $v$ but also on the parameters of the equilibrium point $\{h_\text{opt}(V,V),V\}$ the attaining of which is the goal of the driver actions. Naturally, beyond the perfect rationality approximation the car dynamics cannot be describe in the frameworks of Newtonian mechanics \cite{me2}.   

\subsubsection*{The case of fuzzy rationality}

The perfect rationality cannot be implemented in the reality because of the limit capacity of human cognition. As far as the system at hand is concerned, this limitation manifests itself in the fact that the operator is not able to order states of the system motion by their preference when they are close to one another in quality. In this case the rational behavior becomes physically impossible for the operator and model~\eqref{Int:1} cannot pretend to describe the system dynamics. To tackle this problem let us note the following before modifying the governing equations~\eqref{Int:1}. 

Pursuing two individual succeeding goals can be singled out in the operator actions. The first one is to halt the system motion by driving it to any point $g_t\in\mathbb{L}_\text{pe}$ of the partial equilibrium locus because it could be tough governing fast motion of the system affected not only by the operator intentions but also the physical regularities. The second one is to drive the system towards the equilibrium position $\{0,0\}$, for example, within close proximity to  $\mathbb{L}_\text{pe}$. As a result the mechanisms governing the system motion along the $x$- and $y$-axes are different; let us discuss them separately.  

In the chosen coordinate frame the curve $\mathbb{L}_\text{pe}$ is tangent to the $x$-axis at the equilibrium point $\{0,0\}$, thereby, to simplify the further constructions we may confine our consideration to its certain neighborhood and regard the partial equilibrium locus $\mathbb{L}_\text{pe}$ as the $x$-axis. Outside the partial equilibrium locus the system state varies in time under any action taken by the operator. However, if after driving the system to some point $g_t\in\mathbb{L}_\text{pe}$ the operator fixes the variable $y$, there will be no ``forces'' causing the system motion along  $\mathbb{L}_\text{pe}$; any point of  $\mathbb{L}_\text{pe}$ is steady state. It enables us to approximate the ``force'' $F_x(x,y)$ by a linear function $F_x(x,y) = y\cdot f(x)$ and regard the cofactor $f(x)$ to be mainly determined by the system mechanics. In other words, the fact that the operator behavior is not perfect seems not to influence substantially the rate of system motion along the $x$-axis and, thus, there is no necessity to modify equation~\eqref{Int:1x}. Moreover, in the case under consideration the function $f(x)$ can be regarded as some constant $f(x)=f$ without loss of generality.

The situation is just opposite with respect to the motion along the $y$-axis; the operator actions are to affect it directly. The operator is able to take any reasonable course of actions in order to drive the system towards the partial equilibrium locus and after getting some point $g_t\in\mathbb{L}_\text{pe}$ he can just freeze the system motion along the $y$-axis to halt the system motion as a whole. The point $g_t$ is not necessary to be the point $g\in\mathbb{L}_\text{pe}$ that the operator intended to get initially because reaching any point at the partial equilibrium locus is acceptable to halt the system motion. The point $g$ in turn is not mandatory to coincide with the equilibrium one because of the limit capacity of operator cognition. At the next step of governing the system dynamics the operator has no necessity to be in ``hurry''; now it is possible for him to draw a decision on taking actions for reaching the currently desired point $g$ during a relatively long time interval. In pursuing the latter goal the operator can drive the system either in close vicinity to the partial equilibrium locus $\mathbb{L}_\text{pr}$ or deviating the system from $\mathbb{L}_\text{pr}$ considerably to enable fast motion.

Therefore, to go beyond the frameworks of the perfect rationality in constructing a model for the system motion the following should be taken into account. First, the characteristic time scale $\tau_\text{tr}$ of system dynamics in close vicinity to the partial equilibrium locus $\mathbb{L}_\text{pe}$ must exceed essentially the corresponding time scale far from it, i.e., the inequality $\tau_\text{tr}\gg\tau$ should hold. 

Second, the cognition limitations make it impossible for the operator to locate precisely not only the equilibrium point $\{0,0\}$ at $\mathbb{L}_\text{pe}$ but also the position of the partial equilibrium locus $\mathbb{L}_\text{pe}$ itself. In order to specify this uncertainty let us introduce the perception thresholds, $\{\theta_x,\theta_y\}$, that characterize the dimensions of the neighborhood $\mathbb{Q}_\text{tr}$ of the equilibrium point $\{0,0\}$ withing which this point as well as the corresponding fragment of $\mathbb{L}_\text{pe}$ can be located by the operator with high probability. Since the control over the variable $y$ is of prime priority in governing the system dynamics the threshold $\theta_y$ can be treated as a small parameter. Therefore the region $\mathbb{Q}_\text{tr}$ is actually a some narrow neighborhood of the partial equilibrium locus $\mathbb{L}_\text{pe}$ or, more rigorously, its certain fragment containing the equilibrium point. Any point of  $\mathbb{Q}_\text{tr}$ is regarded by the operator as equilibrium with high probability. 

Third, the point $g\in \mathbb{L}_\text{pe}$ characterizing the course of actions chosen by the operator at a given moment of time is not fixed, it can migrate inside the region  $\mathbb{Q}_\text{tr}$ as time goes on. The movement of this point has to be rather irregular until it remains inside the domain $|g|\lesssim \theta_x$.        

The three features enable us to claim that the standard concept of stability is inapplicable to the system motion near the partial equilibrium locus $\mathbb{L}_\text{pe}$. Indeed, although the system has not reached the desired equilibrium point $\{0,0\}$ the operator freezes its motion near some other point $g_t\in\mathbb{L}_\text{pe}$ and then keeps the system near $\mathbb{L}_\text{pe}$ until he makes a decision about driving the system towards $\{0,0\}$. So, before making this decision the system motion near $\mathbb{L}_\text{pe}$ looks like stable fluctuations near $\mathbb{L}_\text{pe}$, after that it has to be classified as unstable.

Let us discuss a model based on \eqref{Int:1} that captures the key aspects of such operator behavior. The operator chooses a point $g$ on the $x$-axis to which he is going to drive the system. While his control over the system motion is active the system dynamics is governed by the equations 
\begin{subequations}\label{FReq:1}
\begin{align}
\label{FReq:1x}
  \tau\frac{dx}{dt} &= f\cdot y\,,
\\   
\label{FReq:1y}
  \tau\frac{dy}{dt} &= \Omega\cdot F(x-g,y)\,
\end{align}
with the cofactor $\Omega$ equal to unity, $\Omega = 1$. Here the subscript $y$ is omitted at the ``force'' $F(x-g,y)$.  The two equations actually describe the system dynamics outside the region $\mathbb{Q}_\text{tr}$.  When, roughly speaking, the system enters the region  $\mathbb{Q}_\text{tr}$ the operator regarding its any point as an acceptable destination just freezes the system motion along the $y$-axis to such a degree that real variations in the variable $y$ become imperceptible to him and, thus, are not controllable. This action is described by the stepwise transition 
\begin{equation}
\label{FReq:Om1}
   \{\Omega =1\} \Rightarrow \{\Omega =\Delta_r\} \text{\quad with a probability rate}\quad \frac1{\tau} P\left(\frac{x}{\theta_x},\frac{y}{\theta_y}\right)\,.
\end{equation}   
Here $\Delta_r=\Delta_r(t)$ is some small random value, $|\Delta_r|\ll1$, which, in addition, can change in time also in an uncontrollable way,  $P(z_x,z_y)\leq1$ is a certain function of two arguments such that 
\begin{equation*}
\begin{array}{lll}
    P(z_x,z_y) \approx 1  &\quad &  \text{for $|z_x| \lesssim 1$ \textit{and} $|z_y| \lesssim 1$}\,,\\[0.5em]
    P(z_x,z_y) \ll 1      & &  \text{for $|z_x| \gtrsim 1$ \textit{or} $|z_y| \gtrsim 1$}\,.
\end{array}
\end{equation*}
When the system leaves the region $\mathbb{Q}_\text{tr}$ under the actions of uncontrollable factors or the operator gets a decision to correct the system location along the $x$-axis he resumes governing the system dynamics, which is represented as the stepwise transition
\begin{equation}
\label{FReq:Om2}
   \{\Omega =\Delta_r\}  \Rightarrow \{\Omega =1\}  \text{\quad with the probability rate}\quad  \frac1{\tau}\left[1- P\left(\frac{x}{\theta_x},\frac{y}{\theta_y}\right)\right]\,.
\end{equation}
Finally the given model should be completed by an equation describing the operator perception of the desired destination point inside  $\mathbb{Q}_\text{tr}$. In the present paper we write it in a symbolic form
\begin{equation} \label{FReq:G}
   \frac{dg}{dt} = \widehat{R}(x,y,t,g|\theta_x,\theta_y)\,,
\end{equation}
\end{subequations}
where the presence of the time $t$ in the list of arguments allows for random factors in the dynamics of the variable $g$.  

It should be noted that the formulated model of the system dynamics enables us to specify some general features of the ``force'' $F(x-g,y)$.  In fact, the existence of the partial equilibrium locus $\mathbb{L}_\text{pe}$ has allowed us to single out two stimuli in governing the system motion which determine the operator actions. They may be reformulated as follows. The goal of the first one is to keep up the variable $y$ in close proximity to the partial equilibrium locus $\mathbb{L}_\text{pe}$ in order to depress the fast system motion. This stimulus can be quantified in terms of the variable $y$. As result, the component $F_{\text{I}}(x-g,y)$ of the ``force'' $F(x-g,y)$ caused the first stimulus may be written as $F_{\text{I}}(x-g,y)=-\sigma y$, where $\sigma>0$ is some kinetic coefficient. The second one is related to the operator actions of driving the system towards the desired  point $g\in \mathbb{L}_\text{pe}$. The corresponding component $F_{\text{II}}(x-g,y)$ does not change its singe in crossing the $x$-axis and in the simplest case may be represented in the form $F_{\text{II}}(x-g,y)=-\beta (x-g)$, where $\beta>0$ is another kinetic coefficient. Therefore the expression
\begin{equation}\label{OscMod1}
   F(x-g,y) = -\beta (y-b) - \sigma y
\end{equation}
is the simplest ansatz catching the basic features of such systems. The form of the governing equations~\eqref{FReq:1} within approximation~\eqref{OscMod1} allows us call it the oscillator with dynamical traps.  

Following \cite{AP} the time moments when the operator suspends or resumes the control over the system motion will be referred to as action points. Besides, the neighborhood $\mathbb{Q}_\text{tr}$ of the partial equilibrium locus $\mathbb{L}_\text{pe}$ will be called the region of dynamical traps because after transition~\eqref{FReq:Om1} the system can reside inside it for a long time. It should be pointed out that a similar notion of dynamical traps was also introduced  for relaxation oscillations in systems with singular kinetic coefficients \cite{medtr} and congested traffic flow \cite{TFDT}. Besides, the concept of dynamical traps is met in describing Hamiltonian systems with complex dynamics (for a review see, e.g., \cite{Zas}) that denote some regions in the corresponding phase space with anomalously long residence time, however, the nature of the latter traps is different. 

The stated concept of human behavior combing the principles of the perfect rationality and the limit capacity of human cognition in ordering possible actions, events, etc. by their preference and as a results, treating some of them as equivalent will be called the fuzzy rationality. In the case under consideration the fuzzy rationality reflects itself in two effects, the stagnation of the system motion in the region of dynamical traps and probabilistic nature of the system dynamics in this region.

\subsubsection*{Continuous model for the stagnation effect caused by dynamical traps}

In the present paper the main attention is focused on the complexity of system motion, e.g., the corresponding phase portraits, that is caused by the dynamical traps on their own, i.e., dynamical traps treated as sources of the motion stagnation. Previously \cite{me3,me4} it was demonstrated that the cumulative effect of the motion stagnation in the trap region and additive noise can cause nonequilibrium phase transitions. Moreover, for one oscillator with dynamical traps there exists Lyapunov's function, so in this case the role of noise is constructive, i.e., for the corresponding phase transition to arise the noise intensity should exceed some threshold \cite{MTh,me5}. For an ensemble of several oscillators with dynamical traps we have failed to find the corresponding Lyapunov's function and the preliminary results of numerical simulation \cite{MTh} enabled us to pose a hypothesis that the system stagnation in the region of dynamical traps on its own can induce nonequilibrium phase transitions and formation of complex spatial and temporal patterns. In other words, even without noise effects, in multi-element systems with dynamical traps there should be emergent phenomena of a new type caused by the mismatch between actions of different operators with fuzzy rationality.

\begin{figure}
\begin{center}
\includegraphics[width=1\columnwidth]{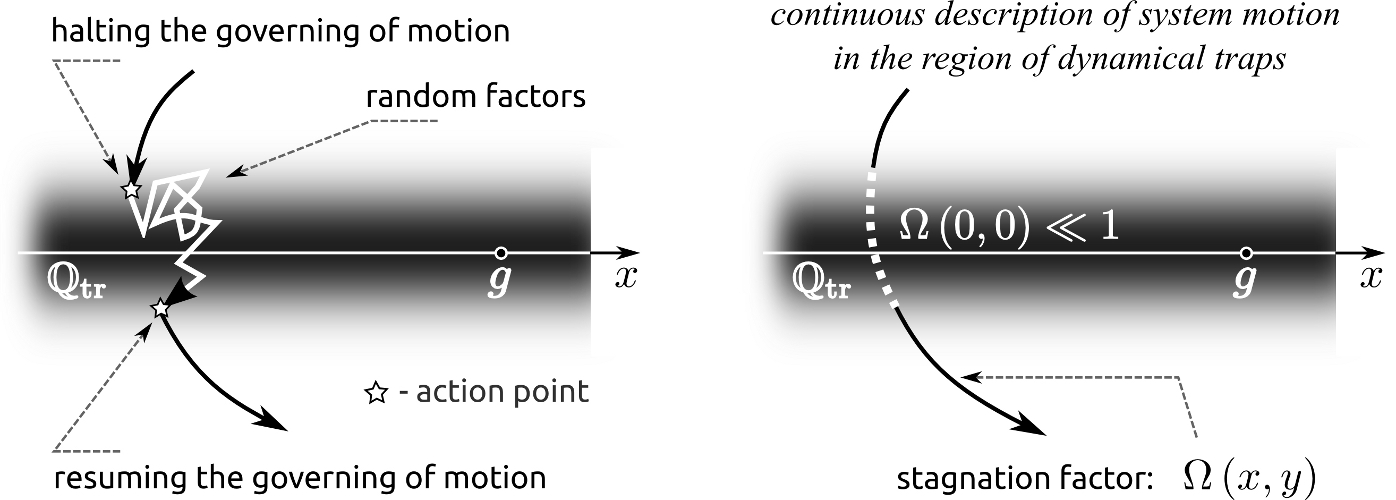}
\end{center}
\caption{The \textit{left fragment} illustrates the characteristic sequence of events in going through the region of dynamical traps $\mathbb{Q}_\text{tr}$, namely, halting the governing of the system motion in entering $\mathbb{Q}_\text{tr}$, then, random motion of the system in $\mathbb{Q}_\text{tr}$, and, finally, resuming the governing of motion due to the system leaving the region $\mathbb{Q}_\text{tr}$ or the operator decision of driving the system towards the desired position $g$. The \textit{right fragment} represents the corresponding effective continuous description of regular system motion in the region $\mathbb{Q}_\text{tr}$ based on introduction of the stagnation factor $\Omega(x,y)$ being continuous function of its arguments meeting the inequality $\Omega(0,0)\ll1$.}
\label{Fcm}
\end{figure}

Therefore, in what follows we will analyze a model that mimics the effects of dynamical traps in the frameworks of regular system motion with stagnation in the region $\mathbb{Q}_\text{tr}$. It is illustrated in Fig.~\ref{Fcm}, where the left fragment depicts the characteristic three events occurring when the system goes through the region of dynamical traps $\mathbb{Q}_\text{tr}$. Namely, halting the system motion when entering $\mathbb{Q}_\text{tr}$, the motion in $\mathbb{Q}_\text{tr}$ under influence of random factors, and, resuming the governing of motion. The latter can be caused by two factors, the system leaving the region $\mathbb{Q}_\text{tr}$ or the operator just getting the decision of driving the system towards the desired point $g\in\mathbb{L}_\text{pe}$. Both of them determine some cumulative mean lifetime $\tau_\text{tr}\gg\tau$ of the system inside  $\mathbb{Q}_\text{tr}$. The right fragment of Fig.~\ref{Fcm} exhibits an effective model mimicking this behavior, at least, semiquantitatively.  It assumes the system motion along the $y$-axis to be governed by the regular force $\Omega(x,y)F(x-g,y)$, where the continuous function $\Omega(x,y)$, the stagnation factor, takes a small value $\Delta \sim \tau/\tau_{tr}\ll1$ at the central points of the region  $\mathbb{Q}_\text{tr}$, in other words, $\Omega(0,0)=\Delta$. As the point $\{x,y\}$ goes away from the central points of  $\mathbb{Q}_\text{tr}$ and leaves it the stagnation factor exhibits gradual growth up to unity, i.e., $\Omega(x,y)\approx 1$ for $|x|\gtrsim \theta_x$ or $|y|\gtrsim \theta_y$. Let us make use of the following ansatz
\begin{equation}\label{CA:1add}
         \Omega \left(x,y\right) = 
			      \frac{\Delta + \left(x/\theta_x\right)^2 + \left(y/\theta_y\right)^2}
                  {1 + \left(x/\theta_x\right)^2 + \left(y/\theta_y\right)^2} \,.
\end{equation}
In addition, in the frameworks of the  given model irregular variations of the final destination point $g$ in the ``mind'' of the operator should be also ignored. Within this approximation model~\eqref{FReq:1} reads
\begin{subequations}\label{FReq:2}
\begin{align}
\label{FReq:2x}
  \tau\frac{dx}{dt} &= f\cdot y\,,
\\   
\label{FReq:2y}
  \tau\frac{dy}{dt} &= \Omega(x,y)\cdot F(x,y)\,.
\end{align}
\end{subequations}
The following constructions for multi-element systems will use these governing equations~\eqref{FReq:2} as the starting point.

\section{Lazy bead model}\label{sec:LBM}

\begin{figure}[t]
\begin{center}
  \includegraphics[width=1.0\textwidth]{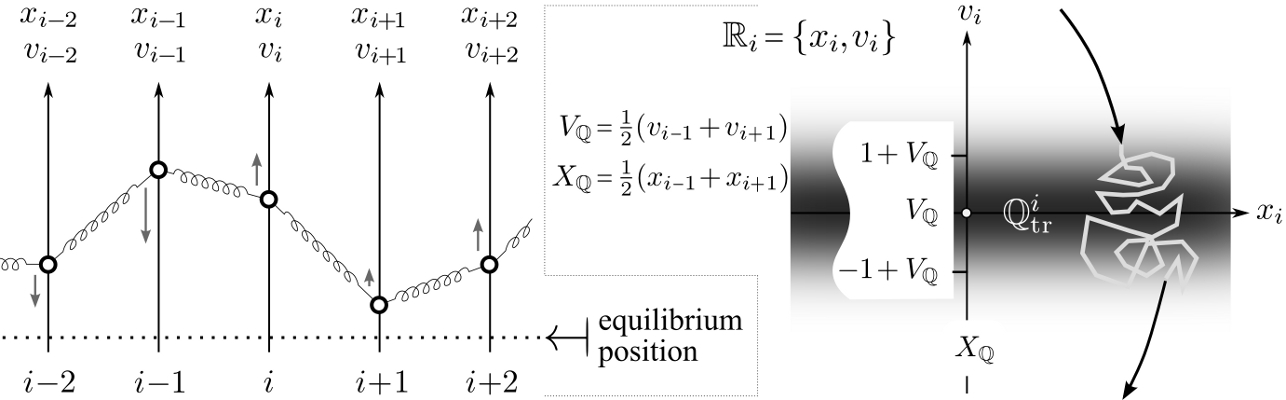}
\end{center}
\caption{The chain of $N$ beads under consideration and the structure of their individual phase space $\mathbb{R}_i=\{x_i,v_i\}$ ($i=1,2,\ldots,N$). The formal initial $i=0$ and terminal $i=N+1$ beads are assumed to be fixed, specifying the equilibrium bead position.}
\label{F2}
\end{figure}

Keeping in mind the aforesaid let us consider a chain of $N$ ``lazy'' beads (Fig.~\ref{F2}). Each of these beads can move in the vertical direction and its dynamics is described in terms of the deviation $x_i(t)$ from the equilibrium position and the motion velocity $v_i(t)=dx_i/dt$ depending on time $t$, here the bead index $i$ runs from $1$ to $N$.  The equilibrium position $x_i=0$ is specified assuming the formal initial ($i=0$) and terminal ($i=N+1$) beads to be fixed. Each bead $i$ ``wishes'' to get the ``optimal'' middle position with respect to its nearest neighbors. So one of the stimuli for it to accelerate or decelerate is the difference 
\begin{equation*}
  \eta_i = x_i -\frac12(x_{i-1} + x_{i+1})
\end{equation*}
provided its relative velocity
\begin{equation*}
 \vartheta_i = v_i -\frac12(v_{i-1} + v_{i+1})
\end{equation*}
with respect to the pair of the nearest beads is sufficiently low. Otherwise, especially if bead $i$ is currently located near the optimal position, it has to eliminate the relative velocity $\vartheta_i$, representing the other stimulus for bead $i$ to change its state of motion. The model to be formulated below combines both of these stimuli within one cumulative impetus $\propto(\eta_i + \sigma\vartheta_i)$, where $\sigma$ is the relative weight of the second stimulus. Actually this ansatz coincides with approximation~\eqref{OscMod1} provided the system variables are measured in units where the kinetic coefficient $\beta=1$. 

When, however, the relative velocity $\vartheta_i$ becomes less then a threshold $\theta$, i.e., $|\vartheta_i|\lesssim\theta$ bead $i$ is not able to recognize its motion with respect to its nearest neighbors. Since a bead cannot ``predict'' the dynamics of its neighbors, it has to regard them as moving uniformly with the current velocities. So from its point of view, under such conditions the current situation cannot become worse, at least, rather fast. In this case bead $i$ just ``allows'' itself to do nothing, i.e., not to change the state of motion and to retard the correction of its relative position. This feature is the reason why such beads are called ``lazy''. Below we will use the dimensionless units in which the perception threshold is equal to unity $\theta =1$ as well as in the later expression for cumulative impetus the required proportionality factor is equal to unity too.   

Under these conditions the equation governing the system dynamics is written in the following form
\begin{equation}
  \label{2}
  \frac{dv_i}{dt}  = - \Omega(\vartheta_i)[\eta_i + \sigma \vartheta_i + \sigma_0 v_i]\,.
\end{equation}
If the cofactor $\Omega(\vartheta_i)$ was equal to unity the given system would be no more then a chain of beads connected by elastic springs characterized by the friction coefficient $\sigma$. The term $\sigma_0v_i$ with the coefficient $\sigma_0\ll1$ that can be treated as a certain viscous friction of the bead motion with respect to the given physical frame has been introduced to prevent the system motion as a whole reaching extremely high velocities. The factor $\Omega(\vartheta_i)$ is due to the effect of dynamical traps and following the general ansatz~\eqref{CA:1add} we write
\begin{equation}
 \label{6}
   \Omega(\vartheta)  = \frac{\Delta+ \vartheta^2}{1+\vartheta^2}\,,
\end{equation}
where, as before, the parameter $\Delta\in[0,1]$ quantifies the intensity of dynamical traps. If the parameter $\triangle = 1$, the dynamical traps do not exist at all, in the opposite case, $\triangle\ll 1$, their influence is pronounced inside the neighborhood $\mathbb{Q}_\text{tr}^i$ of the axis $v_i=(v_{i-1}-v_{i+1})/2$ (the trap region) whose thickness is about unity (Fig.~\ref{F2}). For the terminal fixed beads, $i = 0$ and $i = N+1$, we set 
\begin{equation}\label{7}
 x_0(t)  = 0\,, \qquad x_{N+1}(t)  = 0\,,
\end{equation}
which play the role of ``boundary'' conditions for equation~\eqref{2}.

It should be noted that the emergence phenomena in a similar chain of oscillators with dynamical traps and affected by some additive noise were considered for the first time in papers~\cite{me3,me4}.

\section{Numerical results}\label{sec:NM}

The system of equations~\eqref{2} was solved numerically using the standard explicit Runge-Kutta algorithm of fourth order with fixed time step. Initially all the beads were located at the equilibrium positions $\{x_i|_{t=0} = 0\}$ and perturbations were introduced into the system via ascribing random independent values to their velocities.  The time step $dt$ of numerical integration was chosen in such a way that its decrease or increase by several times have no considerable effects. The system dynamics was found to depend remarkably on the intensity of ``dissipation'' quantified by the parameter $\sigma$. We remind that the parameter $\sigma$ specifies the relative weight of the stimuli to take the middle ``optimal'' position and to eliminate the relative velocity; the larger the parameter $\sigma$, the more significant the latter stimulus. So let us discuss the obtained results for the cases of ``strong'', ``intermediate'', and ``weak'' dissipation individually. 

It should be noted beforehand that, first, all the results of numerical simulation to be presented below were obtained for the dynamical traps of high intensity, namely, for $\Delta = 10^{-4}$. Emergent phenomena in such systems for different values of the dynamical trap intensity as well as the influence of stochastic factors are worthy of individual analysis. Second, the parameter $\sigma_0$ quantifying additional friction introduced to depress extremely high values of the bead velocities $\{v_i\}$ was set equal to $\sigma_0 = 0.01$. Third, in plotting a collection of phase portraits of bead motion, e.g., $\{x_i(t),v_i(t)\}_{i=1}^{N}$, the bead coordinates $\{x_i\}$ are shown with some individual shifts, namely, $x_i\rightarrow x_i+ 50\cdot i$ to simplify the portrait visualization. 

\subsubsection*{``Strong'' dissipation}

\begin{figure}
\begin{center}
  \includegraphics[width=\textwidth]{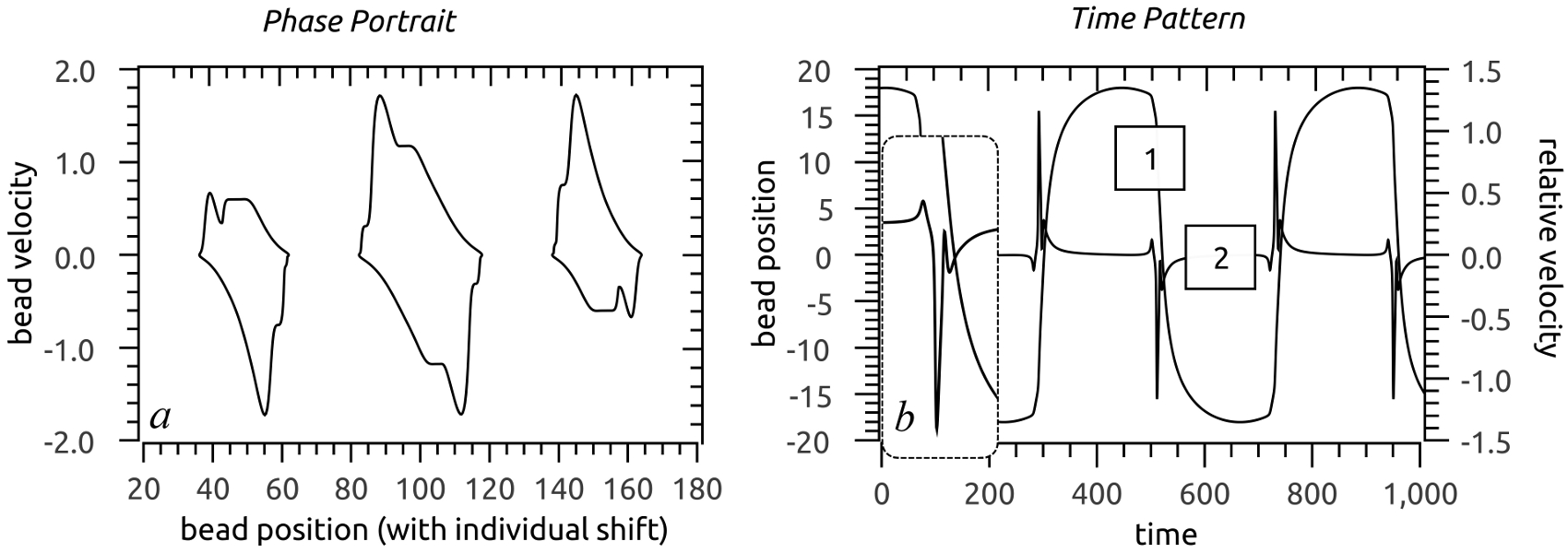}
\end{center}
\caption{The phase portraits of the individual bead motion on the phase plane $\mathbb{R}_{xv}$ for the chain of three beads with ``strong'' dissipation (\textit{a}) and the corresponding time patterns of the middle bead motion (\textit{b}), namely, the time variations in the bead position (1) and the relative velocity (2). In numerical simulation the integration time step $dt=0.01$ was used.}
\label{F3}
\end{figure}

The system dynamics with ``strong'' dissipation is exemplified by numerical data obtained using the kinetic coefficient $\sigma = 3$.  In this case the system instability was detected numerically only for the chains with the number of beads $N\geq3$. Figure~\ref{F3}\textit{a} depicts the found limit cycles of the bead oscillations for the chain with $N=3$. The corresponding time patterns $\{x_2(t)\}$, $\{\vartheta_2(t)\}$ of the middle bead $i=2$ showing time variations of its position and relative velocity are exhibited in Fig.~\ref{F3}\textit{b}. As seen, the periodic motion of these beads looks like relaxation oscillations with the ``slow'' motion fragments matching $\vartheta_i = 0$, i.e., the synchronized motion of neighboring beads. It is worthwhile to note that the given bead periodic motion is not the standard relaxation oscillations related to alternative step-wise transitions between two quasistable states directly specified by system properties. In fact, for the given system only the states $\{\vartheta_i=0\}$ are singled out in properties and  the time patten $\{\vartheta_2(t)\}$ exhibits considerable spike-wise variations only withing the ``fast'' motion fragments, whereas at the other moments of time it is located near the point $\vartheta=0$. In addition, it should be pointed out that the state $\{x_i =0,v_i=0\}$ is metastable, i.e.  stable with respect to small perturbations. Moreover, not all large perturbations in the bead velocities were found numerically to give rise to the limit cycle formation; some of them faded away. However, when the instability was initiated, the steady state oscillations appeared usually after the time interval $T\gtrsim 10^4$ exceeding the period of these oscillations ten-fold. 

\begin{figure}
\begin{center}
  \includegraphics[width=\textwidth]{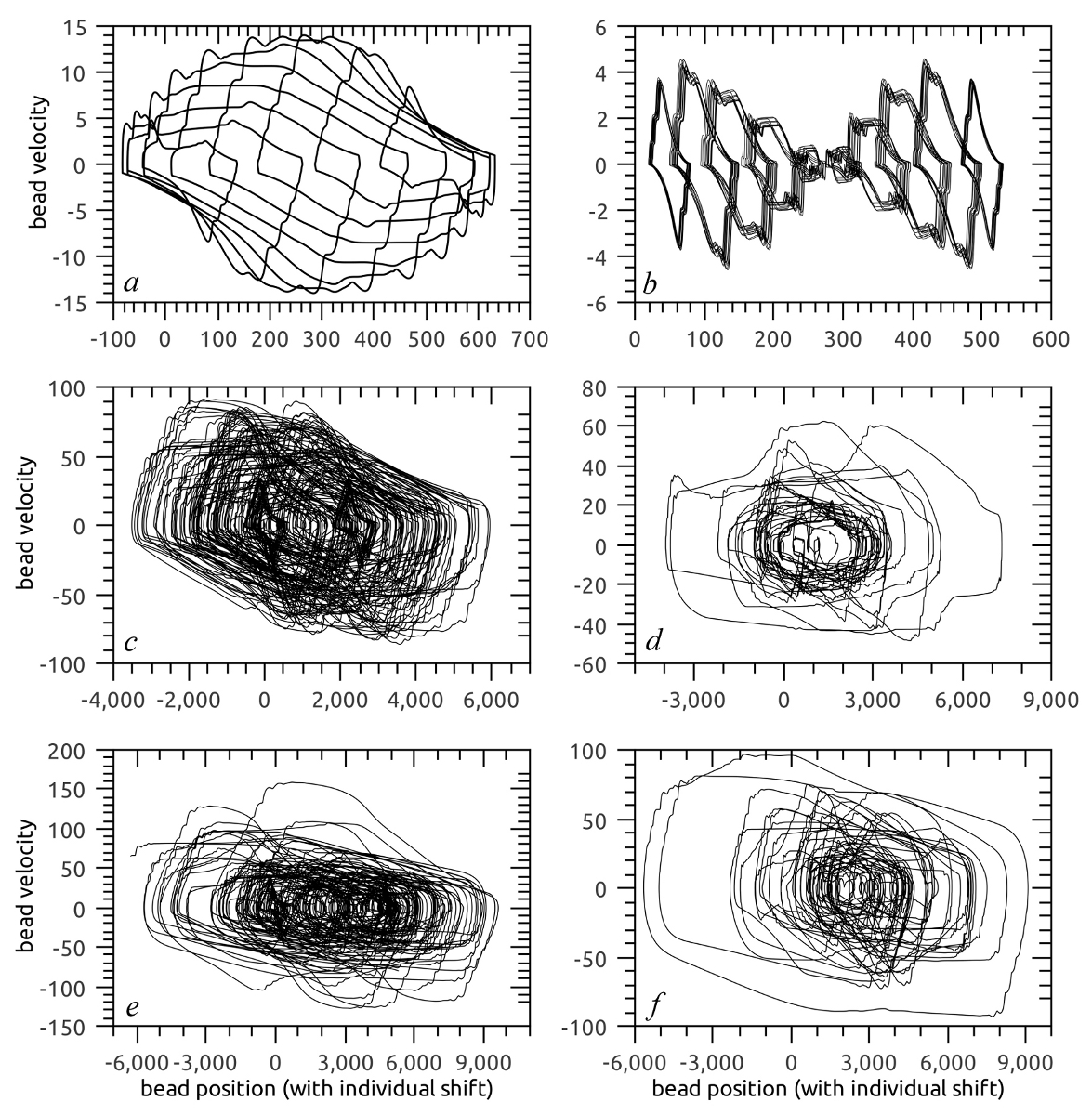}
\end{center}
\caption{Evolution of the phase portraits of individual bead motion on the phase plane $\mathbb{R}_{xv}$ for the chains with ``strong'' dissipation as the number of beads $N$ increases. The frames (\textit{a}) and (\textit{b}) depict the data for $N=10$. The frame (\textit{c}) exhibits the phase portraits of equidistant 10 beads for the chain of 50 beads and a fragment of the corresponding phase trajectory of a middle bead is shown in the frame (\textit{d}). The frames (\textit{e}) and (\textit{f}) demonstrate actually the same data for the chain of 100 beads. In simulation the integration time step $dt= 0.005$ was used. The frames (\textit{d,f}) exhibit trajectories of duration $2\times 10^4$.}
\label{F4}
\end{figure}

For the given bead chains evolution of the phase portraits as the number of beads $N$ increases is illustrated in Fig.~\ref{F4}. While their size is not too large, namely, $N\sim 10$, the regular periodic motion of the beads remains stable, however, various pattens of limit cycles can be formed depending on the initial perturbations. Two found examples are shown in Fig.~\ref{F4}\textit{a,b}. As seen in Fig.~\ref{F4}\textit{b} a limit cycle can have its own complex structure, which does be a property of the system dynamics rather than a numerical artifact; it was verified by decreasing the integration time step by several times. As the number of beads increases the system dynamics becomes irregular (Fig.~\ref{F4}\textit{c,e}), at least, on time scales about $T\sim 10^5$ no periodic bead motion was found for $N\sim 50$. However, the irregularity of individual trajectories seems to grow gradually with the number of beads $N$. The latter feature is demonstrated in Fig.~\ref{F4}\textit{d,f}; the structure of the shown trajectories is visually more regular for the chain of 50 beads in comparison with the chain of 100 beads.  

\subsubsection*{``Intermediate'' dissipation }

\begin{figure}
\begin{center}
  \includegraphics[width=\textwidth]{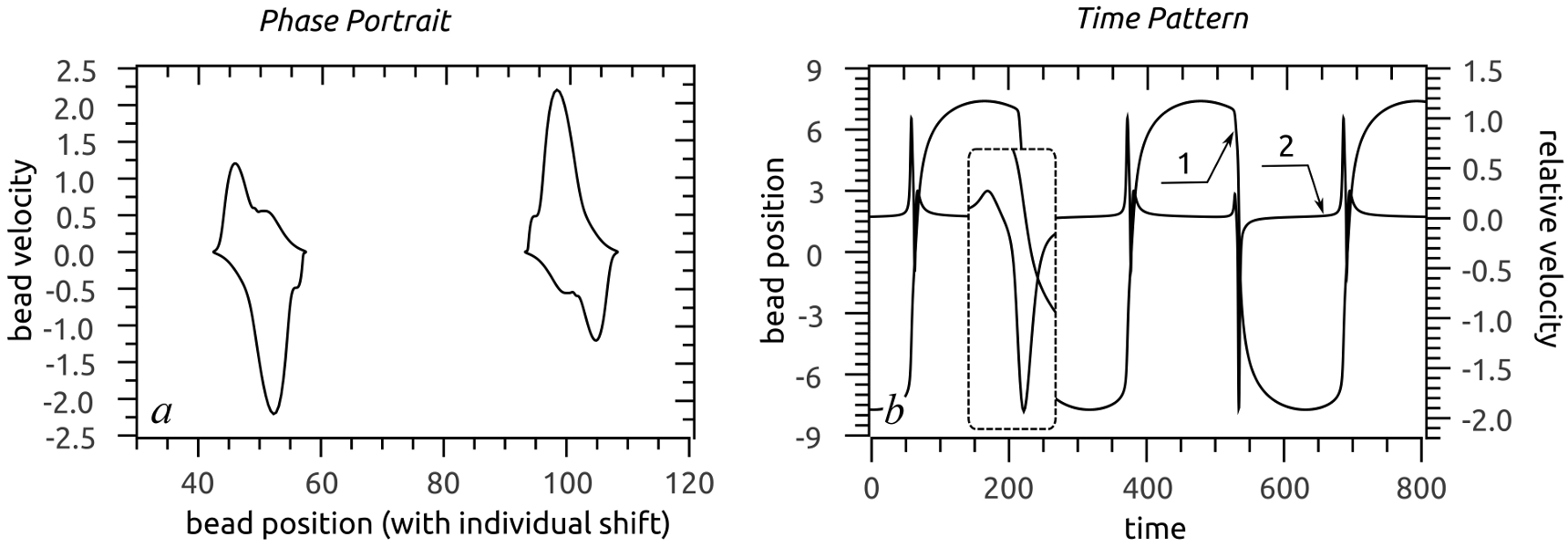}
\end{center}
\caption{The phase portraits of the individual bead motion on the phase plane $\mathbb{R}_{xv}$ for the chain of two beads with ``intermediate'' dissipation (\textit{a}) and the corresponding time patterns of the motion of the first bead (\textit{b}), namely, the time variations in the bead position (1) and the relative velocity (2). In numerical simulation the integration time step $dt=0.01$ was used.}
\label{F5}
\end{figure}

The chains of beads with the kinetic coefficient $\sigma=1$ are treated as characteristic examples of such systems with ``intermediate'' dissipation. In this case the instability was detected in the system of two beads, which is the minimal number of beads when the instability caused by dynamical traps without noise can appear in principle. As noted before, for one oscillator with dynamical traps noise must be present for the instability to arise \cite{me4}.

\begin{figure}
\begin{center}
  \includegraphics[width=\textwidth]{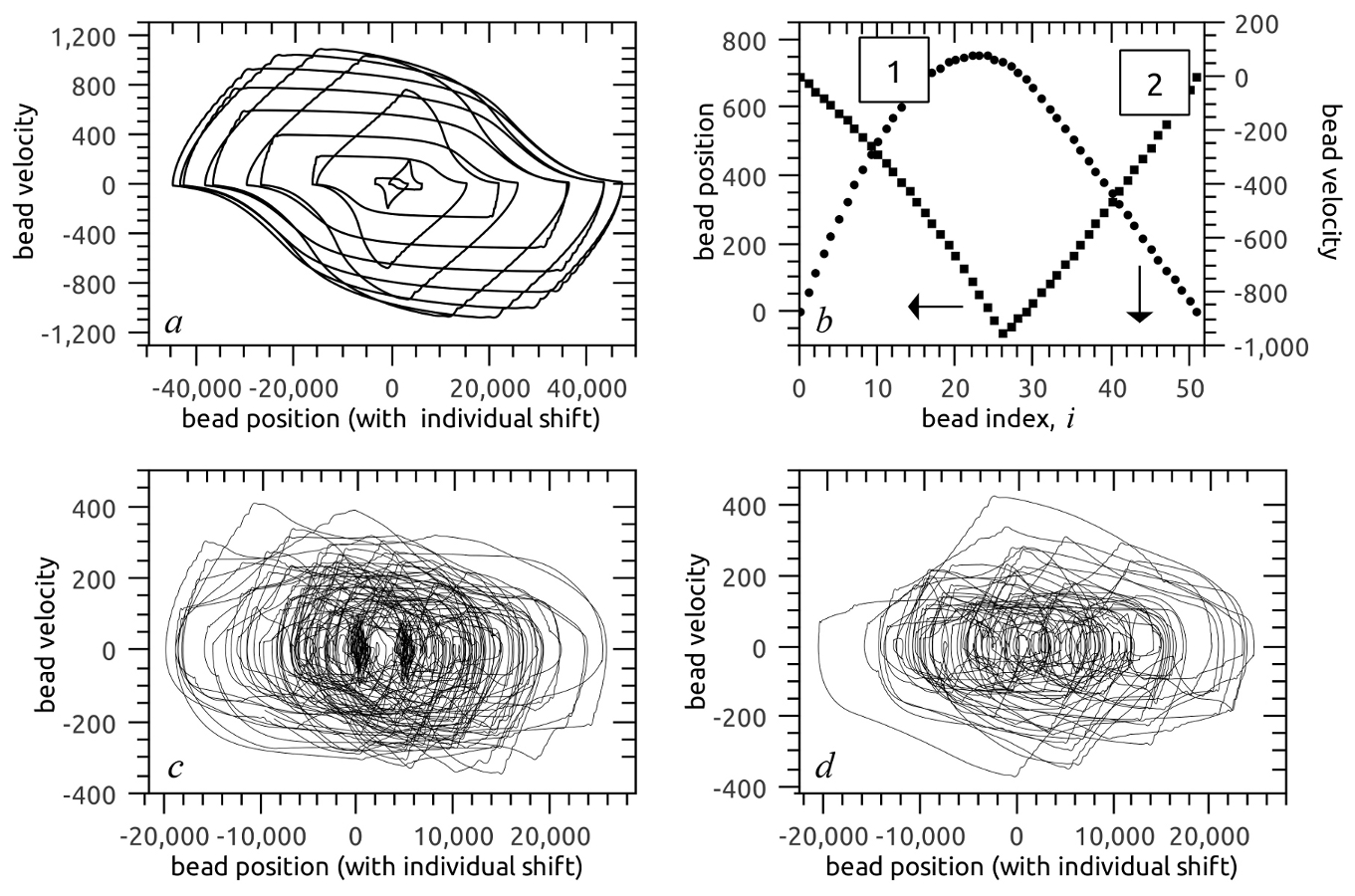}
\end{center}
\caption{The phase portraits of the individual bead motion on the phase plane $\mathbb{R}_{xv}$ for the chains of 50 and 100 beads with ``intermediate'' dissipation, the frames (\textit{a}) and (\textit{c}), respectively. Here 11 equidistant beads are shown. The frame (\textit{d}) depicts the phase portrait of the middle bead motion for the 100 bead chain. For the regular bead motion the frame (\textit{b}) illustrates the characteristic spatial form of the distribution $\{x_i(t)\}_{i=1}^N$ (1) as well as the corresponding distribution $\{v_i(t)\}_{i=1}^N$ (2) of the bead velocities fixed at a certain time moment. Here arrows show the current direction of motion of the points $\{x_i(t)\}$ and $\{v_i(t)\}$ on the phase plane $\mathbb{R}_{xv}$. 
For the irregular motion the shown fragments are of duration of $3\times10^3$ (\textit{c}) and $2\times10^4$  (\textit{d}), the total simulation time was $10^6$. The integration time step $dt=0.005$ was used.}
\label{F6}
\end{figure}

Following the presentation of the previous subsection Figure~\ref{F5} depicts the phase portraits of the two bead chain dynamics (Fig.~\ref{F5}\textit{a}) and the time patterns $\{x_1(t)\}$, $\{\vartheta_1(t)\}$ of the first bead (Fig.~\ref{F5}\textit{b}). As previously, the periodic motion of these beads looks like relaxation oscillations with the anomalous behavior discussed above and again not all the perturbations give rise to the instability onset. However in the case of ``intermediate'' dissipation the regular periodic motion of beads finally arises for the chains of many beads and only one type of limit cycle patterns was detected numerically. In particular, Figure~\ref{F6}\textit{a} exhibits the stable limit cycles developed in the chain of 50 beads; here 11 equidistant beads are shown. The corresponding spatial profile of the bead ensemble treated as a certain ``beaded string''  as well as its velocity profile are exemplified in Fig.~\ref{F6}\textit{b}. Namely, it exhibits the characteristic spatial form of the distribution $\{x_i(t)\}_{i=1}^N$ describing the deviation of the ``beaded string'' from the equilibrium position as well as the distribution of the bead velocities $\{v_i(t)\}_{i=1}^N$ taken at a certain moment of time $t$. We point out that although the found spatial form of the ``beaded string'' oscillations looks like the fundamental mode of elastic string vibration the system dynamics has nothing in common with vibrations of elastic stretched strings. It becomes visible explicitly in the form of the velocity distribution $\{v_i(t)\}_{i=1}^N$ whose dynamics can be represented, at least, qualitatively as the propagation of a certain cusp along the ``beaded string''. As the number $N$ of beads in the chains increases the periodic bead motion either becomes unattainable for the majority of initial perturbations or requires extremely long time to arise; at least, for the chain of 100 beads the system dynamics remained irregular for all the simulations of duration $T\sim 10^6$, which is illustrated in Fig.~\ref{F6}\textit{c,d}.

\subsubsection*{``Weak'' dissipation}
\begin{figure}
\begin{center}
  \includegraphics[width=\textwidth]{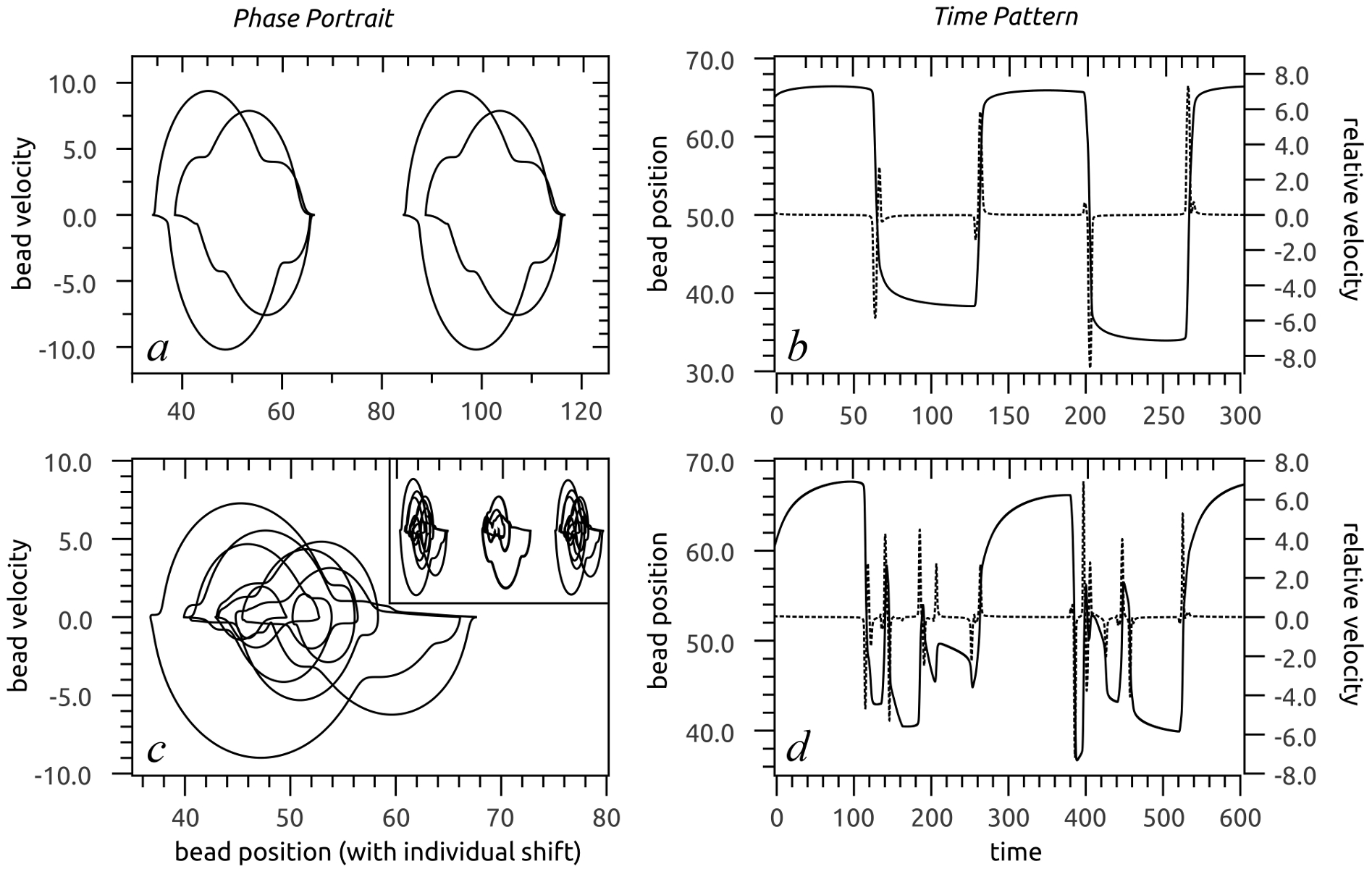}
\end{center}
\caption{The phase portraits of the individual bead motion on the phase plane $\mathbb{R}_{xv}$ for the chains of two and three beads with ``weak'' dissipation (\textit{a,c}) and the corresponding time patterns of the motion of the first and middle beads (\textit{b,d}), namely, the time variations in the bead position (solid line) and the relative velocity (dotted line). The inset in the frame (\textit{c}) depicts the full collection of three phase portraits, whereas its main part exhibits the middle phase portrait in detail. In numerical simulation the integration time step $dt=0.005$ was used.}
\label{F7}
\end{figure}

In the case of ``weak'' dissipation the system dynamics turns out to be more complex in properties, which, by way of example, was analyzed for the bead ensembles with $\sigma = 0.1$. In particular, Figure~\ref{F7} exhibits the phase portraits and the corresponding time patterns for the chains of two and three beads. In both the systems the periodic bead motion is stable; it occurs each time finite amplitude perturbations of the equilibrium state become unstable. The corresponding phase portraits and the time pattern are illustrated in Fig.~\ref{F7}\textit{a,c} and Fig.~\ref{F7}\textit{b,d}, respectively.  However, already for the chain of three beads the limit cycles can have a rather complex form as does the corresponding time patterns of the velocity variations (Fig.~\ref{F7}\textit{c,d}). To make certain that this limit cycle form does be a property of the given system its reproducibility was verified changing the time step in numerical integration or introducing additional small random Langevin forces into equation~\eqref{2}.

\begin{figure}
\begin{center}
  \includegraphics[width=\textwidth]{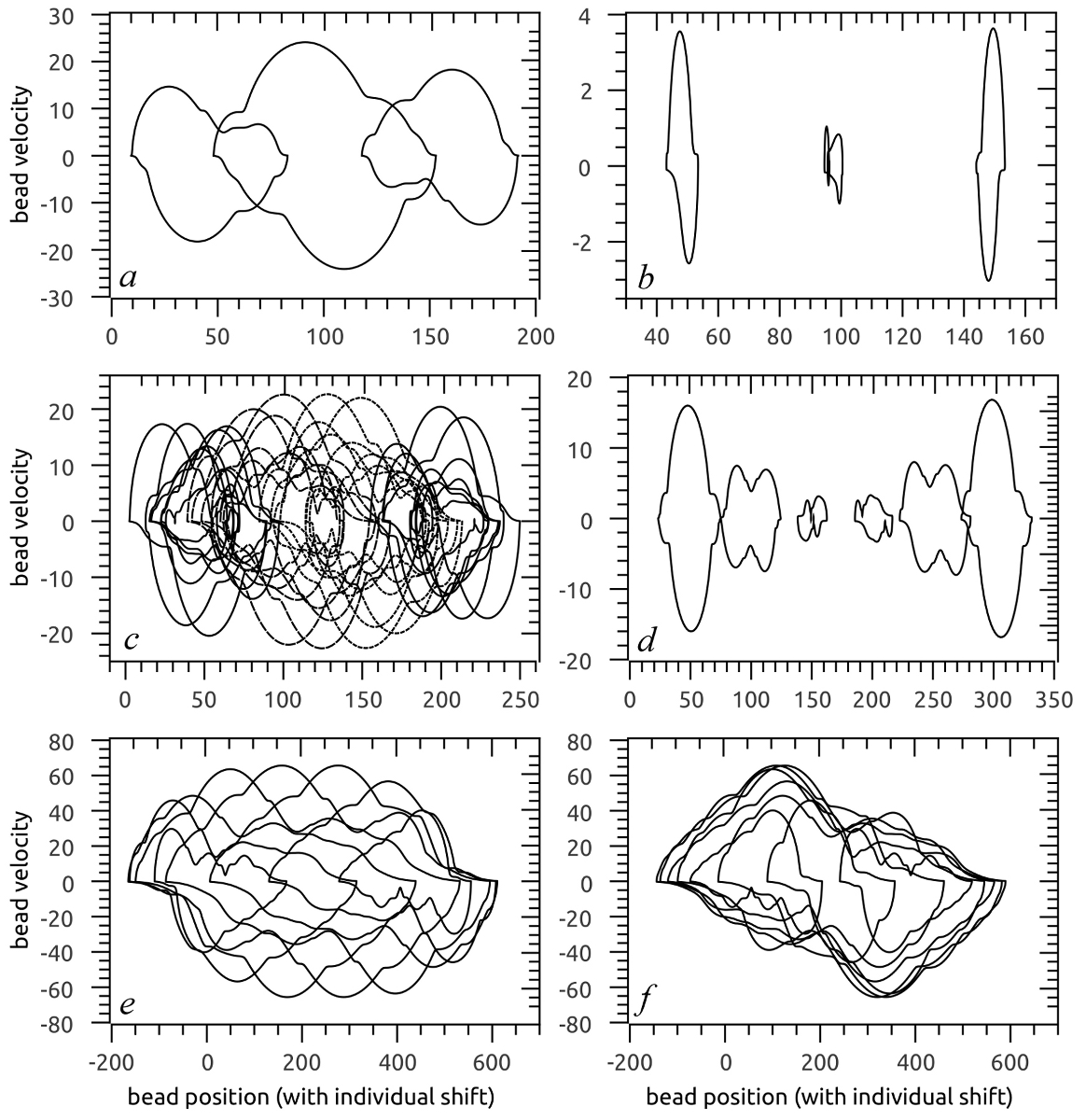}
\end{center}
\caption{The phase portraits of the individual bead motion on the phase plane $\mathbb{R}_{xv}$ for the chains of beads with ``weak'' dissipation. The frames (\textit{a,b}) depict the limit cycles for the three bead ensemble found numerically in addition to one shown in Fig.~\ref{F7}\textit{c}, the frame (\textit{c}) exhibits limit cycles for the four bead chain where the motion trajectories of the second and third beads are plotted out with dashed lines, the plot (\textit{d}) matches the six bead chain, and the frames (\textit{e,f}) show the limit cycle patterns found for the eight bead chain. In numerical simulation the integration time step $dt=0.005$ was used.}
\label{F8}
\end{figure}

As the number $N$ of beads increases the complexity of the system dynamics does not grow gradually, which is exemplified in Fig.~\ref{F8}. For the three bead ensemble two additional types of limit  cycles were fixed (Fig.~\ref{F8}\textit{a,b}) whose structure is rather simple in comparison with one shown in Fig.~\ref{F7}\textit{c}. For the four bead chain (Fig.\ref{F8}\textit{c}) only one type of limit cycles was found numerically; it is similar to one shown in Fig.~\ref{F7}\textit{c} and matches a rather complex periodic motion of individual beads with a relatively large amplitude. To make it clear the limit cycles of the second and third beads are plotted out here with dashed lines. The dynamics of five bead chain is similar in properties. The six bead chain exhibits the opposite behavior illustrated in Fig.~\ref{F8}\textit{d}. The only one stable periodic motion found numerically is of a rather simple geometry and its amplitude is relatively small. The dynamics of seven and eight bead chains is of the same type of complexity, in particular, Figures~\ref{F8}\textit{e,f} plot the found limit cycle patterns for the eight bead ensemble, which can be treated as derivatives of the pattern shown in Fig.~\ref{F8}\textit{a}.

\begin{figure}
\begin{center}
  \includegraphics[width=\textwidth]{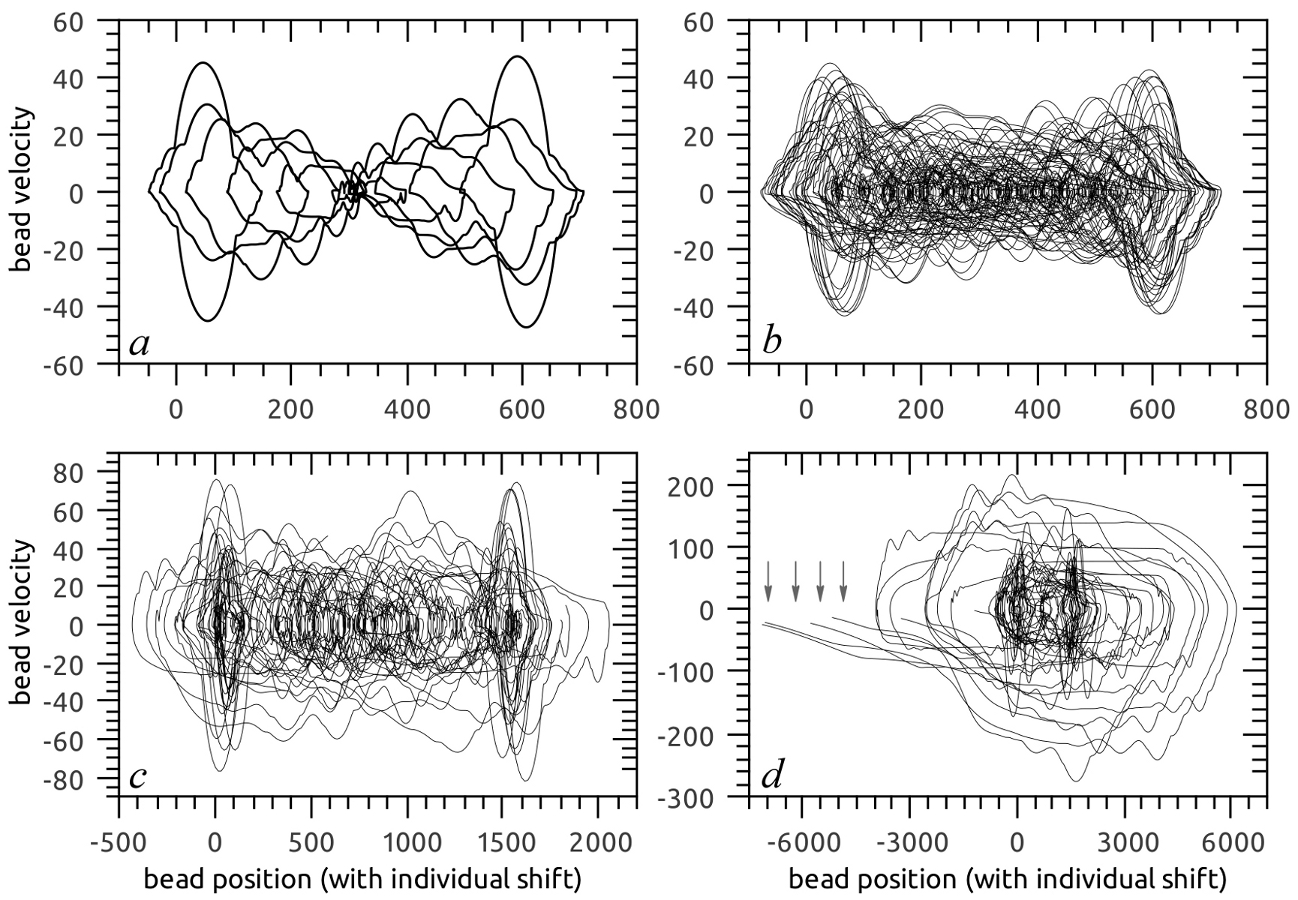}
\end{center}
\caption{The phase portraits of the individual bead motion on the phase plane $\mathbb{R}_{xv}$ for the chains of beads with ``weak'' dissipation. The frame (\textit{a}) depicts a special type of the limit cycle patterns fixed for the 12-bead chain as a rare event, in contrast, the frame (\textit{b}) exhibits an intermediate stage of the instability development observed usually for these chains with the number of beads $N\gtrsim 10$ before limit cycle patterns similar to one shown in Fig.~\ref{F8}\textit{d} appear. Here the bead trajectories of duration of $10^3$ after the simulation time  $T\sim 6\times 10^5$ are presented. The frames (\textit{c,d}) exhibit the transient processes for the 30-bead chain when (\textit{c}) the bead trajectories are located in the vicinity of the limit cycle pattern similar to one shown in the frame (\textit{a}) and at the moment (\textit{d}) corresponding to the explicit formation of the stable periodic motion (pointed out by gray arrows). Here trajectories of duration of $10^3$ are shown after the simulation time $T\sim 7\times 10^5$ and after an additional time interval $\Delta T\sim 2\times 10^4$. In numerical simulation the integration time step $dt=0.003$ and $dt= 0.01$ for the 12-bead and 30-bead chains, respectively, was used.} 
\label{F9}
\end{figure}

As the number of beads increases a new feature of the system dynamics was fixed for the 12-bead ensemble. In addition to the periodic oscillations represented by patterns similar to ones plotted out in Fig.~\ref{F8}\textit{e,f} a limit cycle collection actually of the same form as shown in Fig.~\ref{F8}\textit{b,d} was found (Fig.~\ref{F9}\textit{a}). The given pattern again was verified to be stable with respect to changing the integration time step $dt$ and introducing additional small Langevin forces. Attempts to find a similar periodic motion for the ensembles of 11 or 13 beads were not successful.  Moreover, for the 12-bead ensemble only a few of the generated initial perturbations give rise to it.  A more detailed analysis demonstrated the fact that this type of bead motion is actually an intermediate stage of the instability development for the majority of the generated initial perturbations even for the 12-bead chain. For example, Fig.~\ref{F9}\textit{b} exemplifies the usual geometry of bead trajectories when, on one hand, the initially induced uncorrelated motion of beads has faded away and, on the other hand, the periodic stable motion has not arisen yet. As seen, these trajectories together make up some region looking like the pattern in Fig.~\ref{F9}\textit{a} scatted by some noise. For many-bead ensembles only the stable periodic motion of the type shown in  Fig.~\ref{F8}\textit{f} survives, however, the transient processes of the instability development go through this stage, which is demonstrated in Figs.~\ref{F9}\textit{c,d}. Namely, Figure~\ref{F9}\textit{c} shows the motion trajectories of 7 equidistant beads within the 30-bead ensemble; the shown fragments of duration of $10^3$ match the simulation time $T\sim 7\times10^5$. Figure~\ref{F9}\textit{d} exhibits the bead trajectories of the same system after an additional time interval $\Delta T\sim 2\times 10^4$ when the regular periodic motion of the beads was fixed to start its formation explicitly (in the given figure this moment is pointed out with gray arrows). For large ensembles of beads, for example, the 80-bead chain, the periodic motion was not fixed, at least, on time scales $t\lesssim 10^6$, maybe, because of a fast growth of the required waiting time as the number of beads increases. It poses a question as to whether noise with an extremely small amplitude can cause a stochastic dynamics of such systems. 

\begin{figure}
\begin{center}
  \includegraphics[width=\textwidth]{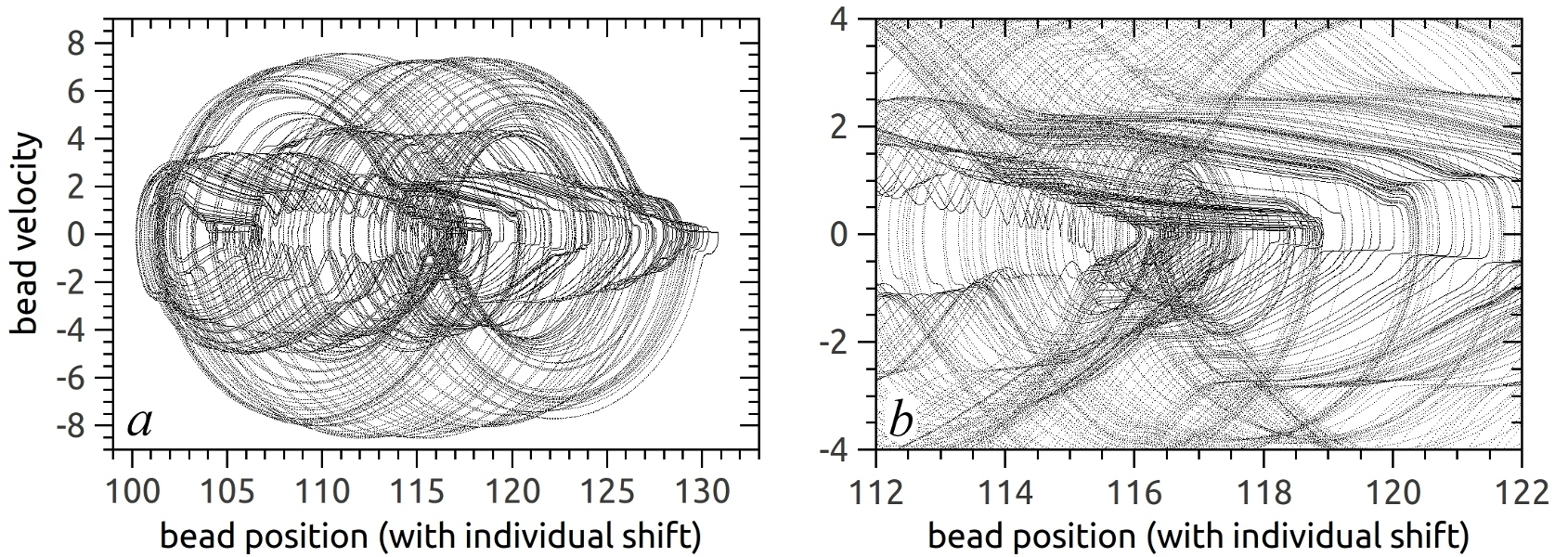}
\end{center}
\caption{The phase portrait of the motion of a middle bead on the phase plane $\mathbb{R}_{xv}$ for the four bead ensemble with $\sigma = 0.03$. The frame (\textit{a}) shows the whole trajectory, whereas the frame (\textit{b}) exhibits its central part. In the plot the bead trajectory is visualized as a sequence of dots separated by the time interval $0.01$.  The shown trajectory fragment is of duration of $2\times 10^3$ and $4\times10^3$ for the plots (\textit{a}) and (\textit{b}), respectively. In numerical simulation the integration time step $dt=5\times 10^{-3}$ was used and the integration time was $T\sim 5\times10^5$.}
\label{F10}
\end{figure}

As the ``dissipation'' parameter $\sigma$ becomes smaller, on one hand, the motion complexity should be met even in the dynamics of ensembles with a few beads. On the other hand, the finite size of such ensembles has to manifest itself in its properties. It is justified by Figure~\ref{F10} demonstrating the phase portrait of the motion of a middle bead in the four bead ensemble with $\sigma = 0.03$. Figure~\ref{F10}\textit{a} depicts the bead trajectory as a whole whereas Figure~\ref{F10}\textit{b} exhibits its central part. Changing the integration time step and the integration time $T$ it was justified that the bound pattern is stable and is not an artifact. As seen, this phase portrait does have a complex multi-scale structure.

\section{Conclusion}\label{sec:Concl} 

A new mechanism of emergent phenomena in social systems governed by cumulative action of human beings and physical regularities has been discussed. It is the fuzzy rationality caused by the bounded capacity of human cognition and manifesting itself in the limited capability of human beings in ordering events, actions, strategies of behavior, etc. according to their preference perfectly. This is most pronounced when, for example, an individual should make a choice between several possible actions similar in quality. As a result, he has to consider them equivalent, thereby, their choice becomes random and practically independent of the real action quality ``hidden'' for the individual. Only in the case where two actions at hand are characterized by a significant difference in quality the choice is determined by the preference relation. When the control over the system dynamics is concerned the fuzzy rationality affects the choice between the ``hidden'' optimal strategy of behavior and actions in its proximity. In this case the optimal strategy becomes unattainable and individuals consider a whole multitude of possible actions ``optimal'' with high probability. As a result, the dynamics of a given system as well as the control by its elements (individuals) is stagnated until the system motion goes rather far from the optimal one, which was expected to induce a system instability. 

To elucidate this concept, first, it has been applied to constructing governing equations for a certain system whose dynamics can be represented as motion of a point $\{x,y\}$ on the phase plane $\mathbb{R}_{xy}$. The system is considered to be governed by some physical regularities and the active behavior of its operator (individual) together, with the contributions of the two factors being of the same significance. This feature is taken into account presuming the existence of a partial equilibrium locus $\mathbb{L}_\text{pe}$ in the phase space $\mathbb{R}_{xy}$, i.e., an one-dimensional set of points such that the system can reside at any one of them infinitely long while the operator suspends its active control over the system dynamics. In other words, the operator can halt the system motion at any point of $\mathbb{L}_\text{pe}$ because when the system gets the partial equilibrium locus $\mathbb{L}_\text{pe}$ there no mechanisms of ``physical'' nature causing its migration along $\mathbb{L}_\text{pe}$; only the motivated behavior of the operator can do this. Outside  $\mathbb{L}_\text{pe}$ the system cannot be halted by any action of the operator; the system is forced to move (on the phase plane $\mathbb{R}_{xy}$) just by the physical regularities. This construction is exemplified in short appealing to the mathematical models for car following. 

Pursuing two individual goals are singled out in the operator actions. The first one is to halt the fast motion by driving the system to any point of the partial equilibrium locus $\mathbb{L}_\text{pe}$. The second one is driving the system towards the desired equilibrium point, for example, in the vicinity of $\mathbb{L}_\text{pe}$. We assume that if the operator behavior were rational strictly, then the motion of the given system would be characterized by some stable stationary point. It should be pointed out that under such conditions self-organized phenomena cannot arise. For the model at hand the implementation of the fuzzy rationality discussed above has been described within the notion of dynamical traps. Namely, the operator is assumed to consider all the points inside a certain neighborhood $\mathbb{Q}_\text{tr}$ of $\mathbb{L}_\text{pe}$ acceptable to be regarded as equilibrium. Therefor, after the system goes into  $\mathbb{Q}_\text{tr}$ called the region of dynamical trap, the operator halts the system motion or his reaction time becomes much longer then the reaction time corresponding to the system motion outside $\mathbb{Q}_\text{tr}$. So, roughly speaking, the fuzzy rationality gives rise to the system stagnation inside the region of dynamical traps rather then induces some instability of the equilibrium point. It has been demonstrated that a simple model of oscillator with dynamical traps catches the general properties of such objects. 

Suspension and resumption of the operator active behavior in governing the system dynamics is a probabilistic process. The present paper, however, analyzes the dynamical trap effect on its own with respect whether it can induce instability and emergent phenomena of a new type in multi-element ensembles. So a continuous deterministic model for the dynamical trap effect was developed. In the frameworks of this model the dynamical trap region is related to anomalous behavior of the corresponding kinetic coefficients inducing the system stagnation in  $\mathbb{Q}_\text{tr}$. So it was possible to expect that in multi-element ensembles the mismatch between actions of different operators should regularly force the system to go away from the region $\mathbb{Q}_\text{tr}$ contained the stationary point, causing an instability of a new type which cannot be met in ``physical'' media. Indeed, this instability, broadly speaking, is due to the partial equilibrium locus $\mathbb{L}_\text{pe}$ being an one-dimensional collection of saddle points rather then stems from some effect making the stationary point unstable. 

To be specific the self-organization of spatio-temporal patterns in the chain of oscillators with dynamical traps, the ``lazy'' bead model, was studied in detail. This model assumes the individual bead behavior to be governed by two stimuli, one of them is to optimize the spatial arrangement of a given bead with respect to its nearest neighbors, the other is to minimize their relative velocities. However, when the relative velocity becomes rather small a bead being ``lazy'' loses motives for active behavior in correcting the current situation because in this case it cannot become worth. This suspension of activity is regarded as the effect of dynamical traps. 

In the frameworks of the ``lazy'' bead model it has been found, in particular, that, first, the dynamical trap effect on its own, i.e., without noise can induce the system instability when the number of beads exceeds some critical value about unity. It should be noted once more that in the case of one oscillator with dynamical traps for the instability to arise noise must be present and its amplitude has to exceed some threshold. 
Second, the complexity of the system dynamics becomes more and more pronounced as the relative weight $\sigma$ of the latter stimulus decreases gradually. For not too small values of $\sigma$ the developed spatio-temporal patterns of the bead motion match periodic motion and in the phase space $\mathbb{R}_{xy}$ they are represented by a collection of limit cycles which, however, can be of a complex form. Nevertheless, when the number of beads becomes large enough, $N\gtrsim 100$, the periodic bead motion was not found numerically, maybe, because the waiting time necessary for these patterns to form becomes too long for these many-element ensembles. It enables us to pose a question as to whether noise with an extremely small amplitude can affect substantially the properties exhibited by systems of many elements with fuzzy rational behavior. Third, in the case of small values of the parameter $\sigma$ it has been demonstrated that even for the ensembles of a few beads the dynamical traps give rise to really irregular system motion and phase portraits with stable multi-scale geometry.

\section*{Acknowledgments}
The work was supported in part by the Competitive Research Funds of the University of Aizu, Project P-4, FY2011, and the Fukushima Prefectural Foundation, Project F-23-1, FY2011.

\end{document}